\documentclass[letterpaper, 10 pt, conference]{ieeeconf}  

\IEEEoverridecommandlockouts                              

\usepackage{graphics} 
\usepackage{epsfig} 
\usepackage{amsmath} 
\usepackage{amssymb}  
\usepackage{hyperref}
\usepackage{mathtools}
\usepackage{xcolor}
\usepackage{bm}
\usepackage{makecell}
\usepackage{tabulary,capt-of}
\usepackage{enumerate}

\usepackage{amsthm,amsfonts}
\newtheorem{theorem}{Theorem}

\newtheorem{definition}{Definition}[section]

\newtheorem{remark}{Remark}[section]

\newcommand{\fref}[1]{Fig.~\ref{#1}}
\newcommand{\sref}[1]{Section~\ref{#1}}

\newcommand{\mat}[1]{\begin{bmatrix}\,#1\,\end{bmatrix}}

\DeclarePairedDelimiter{\norm}{\lVert}{\rVert}

\DeclareMathOperator*{\argmin}{argmin}

\title{\LARGE \bf Control Barrier Function based Quadratic Programs Introduce Undesirable Asymptotically Stable Equilibria}

\author{
Matheus F. Reis, A. Pedro Aguiar and Paulo Tabuada
\\
\thanks{Matheus F. Reis and A. Pedro Aguiar are with the Department of Electrical and Computer Engineering, Faculty of Engineering, University of Porto, Portugal,
        {\tt\scriptsize matheus.reis@fe.up.pt, pedro.aguiar@fe.up.pt}.
Paulo Tabuada is with the Electrical and Computer Engineering Department at the University of California, Los Angeles,
{\tt\scriptsize tabuada@ee.ucla.edu}.
The work of the last author was partially supported by the NSF award 1645824 and by the CONIX Research Center, one of six centers in JUMP, a Semiconductor Research Corporation (SRC) program sponsored by DARPA.}
}

\begin{document}

\maketitle
\thispagestyle{plain}
\pagestyle{plain}

\begin{abstract}
Control Lyapunov functions (CLFs) and control barrier functions (CBFs) have been used to develop provably safe controllers by means of quadratic programs (QPs), guaranteeing safety in the form of trajectory invariance with respect to a given set. In this manuscript, we show that this framework can introduce equilibrium points (particularly at the boundary of the unsafe set) other than the minimum of the Lyapunov function into the closed-loop system. We derive explicit conditions under which these undesired equilibria (which can even appear in the simple case of linear systems with just one convex unsafe set) are asymptotically stable. To address this issue, we propose an extension to the QP-based controller unifying CLFs and CBFs that explicitly avoids undesirable equilibria on the boundary of the safe set. The solution is illustrated in the design of a collision-free controller.
%
%
\end{abstract}

\section{INTRODUCTION}

While the design of asymptotically stabilizing controllers has been extensively studied in control Lyapunov theory \cite{Khalil2002nonlinear}, the design of controllers capable of enforcing invariance of a particular set of states has been the subject of study in the context of control barrier functions (CBFs) \cite{Ames2019}. The concept of barrier functions was initially used in constrained optimization \cite{Forsgren2002} due to their ability to provably establish invariance properties of sets. In \cite{Prajna2004} and \cite{Prajna2006}, barrier certificates were introduced as a tool to formally prove safety of nonlinear and hybrid systems. In the seminal work \cite{Ames2014}, the concept of CBFs was introduced with a novel, less conservative form of barrier constraint, allowing the barrier function value to grow when far away from the boundary of the safe set. This extension allows the system trajectory to approach the boundary of the safe set without ever leaving it. Additionally, \cite{Ames2014} also introduced the idea of unifying CBFs with Control Lyapunov Functions (CLFs) \cite{Sontag1983} through the use of quadratic programs (QPs), effectively combining safety and stabilization requirements in a single, elegant framework suitable for control.
The optimization-based framework introduced by \cite{Ames2014} was followed by a series of related works demonstrating its applicability, such as collision-free control for multi-robot systems \cite{Wang2016}, persistent control for teams of mobile robots \cite{Notomista2018}, bipedal robot walking \cite{Hsu2015}, safe learning of system dynamics \cite{Wang2018}, and adaptive safety using CBFs in the presence of parametric model uncertainty \cite{Taylor2019adaptiveCBFs}.

\begin{figure}
\begin{center}
\includegraphics[width=0.86\columnwidth]{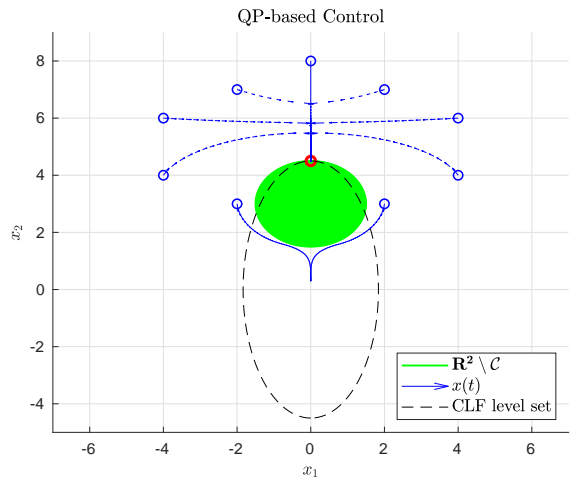}
\end{center}
\vspace{-5mm}
\caption{System trajectories for the QP-based control of the integrator with a circular obstacle. The undesirable asymptotically stable equilibrium is shown in red.}
\label{fig:initial_cond}
\end{figure}

However, the QP-based framework shared across these works suffers from an important limitation. While it guarantees invariance of the system trajectories with respect to the safe set as a hard constraint, it softens the stabilization objective in other to maintain the feasibility of the constrained optimization problem everywhere. 
In this work, we demonstrate using a simple example that this methodology can introduce equilibria other the minimum of the CLF into the closed-loop system, and that these undesirable equilibria can even be asymptotically stable\footnote{An illustration of this phenomena can be seen in \fref{fig:initial_cond} for the integrator system with a convex obstacle. Note that even for the initial condition $x_0=(4,4)$, which is far from the obstacle, the trajectory $x(t)$ does not converge to the origin.}. 
While \cite{Jankovic2018} proposed a modification of the original QP problem seeking to guarantee that the CLF is strictly negative when the barrier constraint is inactive, it still does not address the problem of existence of other types of asymptotically stable equilibria other then the CLF minimum, as we demonstrate in this manuscript. In \cite{Pio2020}, a smooth control law for safety and stabilization was proposed using a different method based on finding and combining different weighted centroids of the feasible control set. However, it remains unclear if this approach actually addresses the problem of undesired equilibria.

%

This paper adds to the literature in the following important ways. First, it demonstrates, both theoretically and by means of numerical simulations, that the QP-based controller with CLF-CBF constraints proposed by \cite{Ames2014} introduces undesired equilibria other than the CLF minimum into the resulting closed-loop system. Secondly, a {\it sufficient} condition under which the resulting undesired equilibria are asymptotically stable are explicitly derived for the integrator system. Finally, we propose an extension to the QP-based controller unifying CLFs and CBFs that explicitly avoids undesirable equilibria on the boundary of the safe set. The solution is illustrated in the design of a collision-free controller.

{
{\bf Notation.}
The operator $\nabla \!: C^1(\mathbb{R}^n) \rightarrow \mathbb{R}^n$ is defined as the {\it gradient} $\frac{\partial}{\partial x}$ of a scalar-valued differentiable function with respect to $x$. We use the notation $L_f V$ to denote the Lie derivative of a differentiable function $V : \mathbb{R}^n \rightarrow \mathbb{R}$ along the vector fields $f$ and $g$, that is, $L_f V = \nabla V^\mathsf{T} f \in \mathbb{R}$ and $L_g V = \nabla V^\mathsf{T} g \in \mathbb{R}^{1 \times m}$, respectively. 
We denote linear dependency between two vectors $v, w \in \mathbb{R}^n$ by $v \parallel w$. Define the {\it scaled orthogonal projection}
$\mathcal{P}_{v} = \norm{v}^2 I_n - v v^\mathsf{T} \in \mathbb{R}^{n \times n}$
for a vector $v \in \mathbb{R}^n$, which is a scaled version of the matrix represention for the orthogonal projection operator defined over $\mathbb{R}^n$. 
%
It has the following useful properties: (i) $\mathcal{P}_{v} = \mathcal{P}^\mathsf{T}_{v}$ (symmetry), (ii) $\mathcal{P}^2_{v} = \norm{v}^2\mathcal{P}_{v}$; (iii) the spectrum of $\mathcal{P}_{v}$ is composed of 0 and $\norm{v}^2$ with algebraic multiplicity 1 and $n-1$, respectively;
(iv) $\mathcal{P}_{v} \, z = \norm{v}^2 z$ for all $z \in \mathbb{R}^n$ on the projective subspace defined by $v \in \mathbb{R}^n$ (that is, such that $z^\mathsf{T} v = 0$); (v) $\mathcal{P}_{v} \, w = 0$ for all $w \in \mathbb{R}^n$ such that $v \parallel w$.
For a matrix $M \in \mathbb{R}^{n \times m}$, the set $\mathcal{N}(M) = \{ x \in \mathbb{R}^m | M x = 0 \}$ denotes the kernel or nullspace of $M$. The set $\mathcal{SO}(n)$ is the special orthogonal group of dimension $n$, consisting of all orthogonal matrices $M$, i.e., $M^\mathsf{T} M = I_n$ of determinant $1$, and the set $\mathfrak{so}(n)$ consists of the corresponding special orthogonal Lie algebra of $\mathcal{SO}(n)$. The operator $\wedge: \mathbb{R}^{\frac{1}{2}n(n-1)} \rightarrow \mathfrak{so}(n)$ is the skew-symmetric map from the real vector space of dimension $\mathbb{R}^{\frac{1}{2}n(n-1)}$ to the corresponding Lie algebra $\mathfrak{so}(n)$.
%
%
For example, for $n = 3$, $\hat{\omega} \in \mathbb{R}^{3 \times 3}$ represents the usual cross product operator. 
The operator $\mathcal{O}_n: \mathbb{R}^{n} \rightarrow \mathbb{R}^{n \times \frac{n}{2}(n-1)}$ is defined by $\hat{\omega} x = \mathcal{O}_n(x) \omega$. For $n = 3$, using the anticommutativity of the cross product, we have $\hat{\omega} x = -\hat{x} \omega$, and therefore $\mathcal{O}_3(x) = -\hat{x}$.
}

\section{Quadratic Programs for Safety Critical Systems}

%
Consider the nonlinear, control-affine system
\begin{align}
\dot{x} \!=\! f(x) \!+\! g(x)u
\label{eq:nonlinear_system}
\end{align}
where $x \in \mathbb{R}^n$ is the system state and $u \in \mathbb{R}^m$ is the control input. The vector fields $f: \mathbb{R}^n \rightarrow \mathbb{R}^n$, $g: \mathbb{R}^n \rightarrow \mathbb{R}^{n \times m}$ are locally Lipschitz. The notation of $G(x) = g(x) g(x)^\mathsf{T}$ will also prove to be useful.
%

%


\begin{definition}[CLFs]
A positive definite function $V$ is a {\it control Lyapunov function} (CLF) for system \eqref{eq:nonlinear_system} if it satisfies:
\begin{align}
\inf_{u\in \mathbb{R}^m} \left[ L_f V(x) + L_g V(x) u \right] \le -\gamma(V(x)) \nonumber
\end{align}
where $\gamma: \mathbb{R}_{\ge 0} \rightarrow \mathbb{R}_{\ge 0}$ is a class $\mathcal{K}$ function \cite{Khalil2002}.
\end{definition}
This definition means that there exists a set of controls that makes the CLF strictly decreasing everywhere outside of the origin. CBFs can be used to design controllers enforcing stability. The corresponding set of stabilizing controls is
%
\begin{align}
K_{clf}(x) \!=\! \{ u \!\in\! \mathbb{R}^m: L_f V(x) \!+\! L_g V(x) u \le -\gamma(V(x)) \} \,. \nonumber
\end{align}
%


In constrast, safety can be framed in the context of enforcing invariance of a particular set of states.
%
Consider the {\it safe} set $\mathcal{C}$ defined as the superlevel set of a continuously differentiable function $h : \mathcal{D} \subset \mathbb{R}^n \rightarrow \mathbb{R}$, as \cite{Ames2019}
\begin{align}
\mathcal{C} &= \{x \in \mathcal{D} \subset \mathbb{R}^n: h(x) \ge 0\} 
\nonumber \\
\partial \mathcal{C} &= \{x \in \mathcal{D} \subset \mathbb{R}^n: h(x) = 0\} \nonumber \\
\text{int}(\mathcal{C}) &= \{x \in \mathcal{D} \subset \mathbb{R}^n: h(x) > 0\} \,. \nonumber
\end{align}
Let $k(x)$ be a feedback controller such that the closed-loop system 
\begin{align}
\dot{x} = f_{cl}(x) := f(x) + g(x)k(x)
\label{eq:feedback_system}
\end{align}
is locally Lipschitz. Then, for any initial condition $x(0) \in \mathcal{D}$ there exists a maximum interval of existence $I(x(0)) = [0, \tau_{max})$ such that $x(t)$ is the unique solution to \eqref{eq:feedback_system} on $I(x(0))$ (if $f_{cl}$ is forward complete, $\tau_{max} = \infty$).
\begin{definition}[Safety]
\label{safe}
The set $\mathcal{C}$ is forward invariant if, for every $x(0) \in \mathcal{C}$, $x(t) \in \mathcal{C}$ for all $t \in I(x(0))$. System \eqref{eq:feedback_system} is safe with respect to the set $\mathcal{C}$ if $\mathcal{C}$ is forward invariant.
\end{definition} 

\begin{definition}[CBFs]
Let $\mathcal{C} \subset \mathcal{D} \subset \mathbb{R}^n$ be the superlevel set of a continuously differentiable function $h : \mathcal{D} \rightarrow \mathbb{R}$, then $h$ is a {\it Control Barrier Function} (CBF) for \eqref{eq:nonlinear_system} if there exists a locally Lipschitz extended class $\mathcal{K}_{\infty}$ function \footnote{An extended class $\mathcal{K}_{\infty}$ function $\alpha: \mathbb{R} \rightarrow \mathbb{R}$ is strictly increasing with $\alpha(0) \!=\! 0$.} $\alpha$ such that
\begin{align}
\sup_{u\in \mathbb{R}^m} \left[ L_f h(x) + L_g h(x) u \right] \ge -\alpha(h(x)) \quad \forall x \in \mathcal{D} \,. \nonumber
\end{align}
\end{definition}
The definition simply means that a CBF $h(x)$ is only allowed to decrease 
on the interior of the safe set $\text{int}(\mathcal{C})$, but not on its boundary $\partial \mathcal{C}$.
%
The set of control values that render $\mathcal{C}$ forward invariant can be formally defined as
\begin{align}
K_{cbf}(x) \!=\! \{ u \!\in\! \mathbb{R}^m: L_f h(x) \!+\! L_g h(x) u \!+\! \alpha(h(x)) \ge 0\} \,. \nonumber
\end{align}

\subsection{Quadratic Program Formulation}

The minimum-norm controller proposed by \cite{Ames2014} is
\begin{align}
\mat{ k(x)^\mathsf{T} \!\!&\!\! \delta(x) }^\mathsf{T} &\!\!=\!\! \argmin_{(u,w)\in\mathbb{R}^{m+1}} \, \norm{u}^2 + p w^2 \label{eq:QP_control} \\
& s.t. \quad L_f V(x) \!+\! L_g V(x) u \le -\gamma(V(x)) + w \nonumber 
\\
& \quad \quad \,\,\, L_f h(x) + L_g h(x) u \ge -\alpha(h(x)) \nonumber
\end{align}
where $p$ is a positive constant.
The objective is to minimize the norm of the control signal and of an auxiliary {\it relaxation} variable $w$, while satisfying the CLF and CBF constraints. The CBF constraint guarantees that $u \in K_{cbf}(x)$, keeping the system trajectories invariant with respect to the safe set. The relaxation variable in the CLF constraint {\it softens} the stabilization objective, maintaining the feasibility of the QP.
%

%
In the next section, we show that the closed-loop system \eqref{eq:feedback_system} with control $k(x)$ from \eqref{eq:QP_control} has undesirable equilibrium points other than the origin, and that these points can be asymptotically stable.

\section{Analysis of the closed-loop system with QP-based controller}

In this section, we investigate some aspects regarding the existence of equilibrium points on the closed-loop system \eqref{eq:feedback_system} with $k(x)$ given by \eqref{eq:QP_control} and their stability properties.

\subsection{Existence of equilibrium points}

We now present an important result describing the closed-loop equilibria.
The proof is presented in Appendix \ref{proof1}.
%

\begin{theorem}
\label{theorem1}
The set $\mathcal{E}$ of equilibrium points of the closed-loop system resulting from the application of the control law \eqref{eq:QP_control} into \eqref{eq:nonlinear_system} is given by
\begin{align}
\mathcal{E} = \{ 0 \} \cup \mathcal{E}_{\text{int}} \cup \mathcal{E}_{\partial \mathcal{C}} \nonumber
\end{align}
where $0 \in \mathbb{R}^{n}$ is the origin of the state space and
\begin{align}
\mathcal{E}_{\text{int}} &\!=\! \Big\{ x \in {\bf \Omega_{\overline{cbf}}^{clf}} \!\setminus\! \{ 0 \} | f(x) \!=\! p \gamma(V(x)) G(x) \nabla V(x) \Big\}
\label{eq:interior_equilibria}
\\
\mathcal{E}_{\partial \mathcal{C}} &\!=\! \Big\{ x \in {\bf \Omega_{cbf}^{clf}} \cap \partial \mathcal{C} \,|\,
\mathcal{N}\!\left( \! \begin{bmatrix}
f(x)^\mathsf{T} \\ 
\nabla V(x)^\mathsf{T} G(x) \\ 
\nabla h(x)^\mathsf{T} G(x)
\end{bmatrix}^{\!\!\mathsf{T}} \right) \!\setminus\! \{ 0 \} \ne \emptyset \Big\}
\label{eq:boundary_equilibria}
\end{align}
where $G(x) = g(x) g(x)^\mathsf{T}$, $\mathcal{E}_{\text{int}}$ is the set of {\bf interior equilibria} and $\mathcal{E}_{\partial \mathcal{C}}$ is the set of {\bf boundary equilibria}.
The set ${\bf \Omega_{\overline{cbf}}^{clf}}$ denotes the states where the CLF constraint in \eqref{eq:QP_control} is active and the CBF constraint is inactive, while ${\bf \Omega_{cbf}^{clf}}$ denotes the states where both CLF and CBF constraints are active:
\begin{align}
{\bf \Omega_{\overline{cbf}}^{clf}} = \Big\{ x \in \mathbb{R}^n : L_f V \!+\! \gamma(V) \ge 0 \,, \nonumber \quad \quad \quad \quad \quad \quad \quad \quad \quad \\
L_g V L_g h^\mathsf{T} ( \mathcal{L}_f V \!+\! \gamma(V) ) < ( L_f h \!+\! \alpha(h) ) ( p^{-1} \!+\! \norm{\mathcal{L}_g V}^2 ) \Big\} \nonumber \\
{\bf \Omega_{cbf}^{clf}} = \Big\{ x \in \mathbb{R}^n : L_g V L_g h^\mathsf{T} \left( \frac{L_f h + \alpha(h)}{L_f V + \gamma(V)} \right) \le \norm{L_g h}^2 \,, \nonumber \\ 
L_g V L_g h^\mathsf{T} \ge \left( \frac{L_f h \!+\! \alpha(h)}{L_f V \!+\! \gamma(V)} \right) (\norm{L_g V}^2 \!+\! p^{-1} ) \Big\} \nonumber
\end{align}
%
%
\end{theorem}


\begin{remark}
\label{remark_boundary}
Regarding the boundary equilibria, in general, the existence of a nontrivial null space for the matrix $\begin{bmatrix}
f(x) \!&\! g(x) \mathcal{L}_g V(x)^\mathsf{T} \!&\! g(x) \mathcal{L}_g h(x)^\mathsf{T}
\end{bmatrix}$ implies collinearity among vectors $f(x)$, $g(x) \mathcal{L}_g V(x)^\mathsf{T}$ and $g(x) \mathcal{L}_g h(x)^\mathsf{T}$. 
For an integrator $\dot{x} = u$ with a convex CLF, 
$\mathcal{E}_{\partial \mathcal{C}}$ is simply 
$$
\mathcal{E}_{\partial \mathcal{C}} = \{ x \in {\bf \Omega_{cbf}^{clf}} \cap \partial \mathcal{C} \,|\, \nabla V(x) \!\parallel\! \nabla h(x) \} \,.
$$
\end{remark}

%

\subsection{Stability of equilibrium points}

It was already shown that, in general, the origin $x = 0$ is not an unique equilibrium point of the closed-loop system \eqref{eq:feedback_system} with $k(x)$ given by \eqref{eq:QP_control}. 
%
In this section, the objective is to estabilish an example showing that some of these equilibria can be asymptotically stable.

\begin{theorem}
\label{theorem2}

Consider the integrator $\dot{x} = u$ with $u = k(x)$ given by \eqref{eq:QP_control} with a convex CLF $V(x)$. An equilibrium point $x^\star \in \mathcal{E}_{\partial \mathcal{C}}$ is asymptotically stable if
\begin{align}
H_{V}(x^\star) - c H_{h}(x^\star) > 0 \nonumber
\end{align}
where $c \in \mathbb{R}_{\ge 0}$ is the constant satisfying $\nabla V(x^\star) = c \nabla h(x^\star)$ and $H_{V}$, $H_{h}$ are the Hessian matrices of the CLF and the CBF, respectively.
%
%
\end{theorem}
%
The proof of this result is presented in Appendix \ref{proof2}.
%
%
%
In \fref{fig:initial_cond}, we illustrate Theorem \ref{theorem2} by means of an example. We use the CLF $V(x) = 0.5 \lambda_1 x_1^2 + 0.5 \lambda_2 x_2^2$, $\lambda_1 > \lambda_2$ and CBF $h(x) = 0.5 \norm{x - x_c}^2 - 0.5 r^2$, whose superlevel set $h(x) = 0$ models the boundary of a circular obstacle with radius $r$ centered on $x_c \in \mathbb{R}^2$. The set of boundary equilibria is given by 
$\mathcal{E}_{\partial \mathcal{C}} = \{ x \in {\bf \Omega_{cbf}^{clf}} \cap \partial \mathcal{C} \,|\, \Lambda x = c (x - x_c) ,\, c \in \mathbb{R}\}$ in this case. System trajectories starting close to the top of the obstacle converge to the asymptotically stable equilibrium point shown in red.
Observing the CLF level set at the equilibrium and the boundary $\partial \mathcal{C}$, note that the local curvature of the CBF 
is smaller than the local curvature of $H_V(x)$ at the equilibrium, as expected from Theorem \ref{theorem2}.


\section{Lyapunov Shaping for QP-based Controllers}
\label{sec:design}

We now seek to design a stabilizing controller $k(x)$ such that the resulting closed-loop system \eqref{eq:feedback_system} does not contain certain types of undesired equilibria. 
%
%
%
%
Consider that a positive-definite, {\it non-radial}
\footnote{
A radial function $r: \mathbb{R}^n \rightarrow \mathbb{R}$ is defined by the property $r(Q x) = r(x)$ for all $Q \in \mathcal{SO}(n)$. That is, $r$ is invariant under rotations around the origin.
}
reference CLF $V_r : \mathbb{R}^n \rightarrow \mathbb{R}$ is given.

\begin{remark}
As an example, the reference CLF could be the quadratic form $V_r(x) = \frac{1}{2} x^\mathsf{T} \Lambda x$, where $\Lambda > 0 \in \mathbb{R}^{n \times n}$ is a diagonal matrix with at least a pair of distinct eigenvalues. 
\end{remark}

Next, define another CLF $V: \mathbb{R}^n \times \mathcal{SO}(n) \rightarrow \mathbb{R}$ as
\begin{align}
V(x,Q) = V_r(Q x) 
\label{eq:rotCLF}
\end{align}
with $Q \in \mathcal{SO}(n)$.
%
The time derivative of $Q$ is
%
\begin{align}
\dot{Q} = Q \, \hat{\omega}
\label{eq:Q_dynamics}
\end{align}
%
where $\omega \in \mathbb{R}^{\frac{1}{2}n(n-1)}$ is a virtual control signal with the dimension of $\mathfrak{so}(n)$.
%
%
%
%
%
Using \eqref{eq:feedback_system} and the properties of the skew-symmetric map $\wedge$, it is possible to show that the time derivative of $V(x,Q)$ is {\it affine} with respect to $u$ and $\omega$.
%
%
%
%


As pointed out by Remark \ref{remark_boundary}, the existence of boundary equilibria is connected to the existence of sets where vectors $f(x)$, $G(x) \nabla V(x,Q)$ and $G(x) \nabla h(x)$ are pairwise collinear. Motivated by this fact, we design a function that measures the proximity of the trajectories to these sets:
\begin{align}
\mathcal{D}(x,Q) 
&\!=\! \frac{1}{2} \nabla V(x,Q)^\mathsf{T} G\! \left( \mathcal{P}_{f} \!+\! \mathcal{P}_{G \nabla h} \right) \!G \nabla V(x,Q)
\label{eq:parallel}
\end{align}
%
\begin{remark}
\label{remark_collinearity}
Using property (v) of the scaled orthogonal projection, note that \eqref{eq:parallel} is zero for all the combinations of collinearity conditions under which $\begin{bmatrix} f(x) \!\!&\!\! G(x) \nabla V(x,Q) \!\!&\!\! G(x) \nabla h(x) \end{bmatrix}$ has a nontrivial null space, which includes both points on $\mathcal{E}_{\partial \mathcal{C}}$ and on $\mathcal{E}_{\text{int}}$. Therefore, $\mathcal{D}(x,Q)$ is zero for all $x \in \mathcal{E}_{\partial \mathcal{C}} \,\cup\, \mathcal{E}_{\text{int}}$, $Q \in \mathcal{SO}(n)$, and strictly positive elsewhere.
\end{remark}

Aiming at avoiding the collinearity conditions imposed by the set $\mathcal{D}(x,Q) = 0$ when the trajectories approach the boundary $\partial \mathcal{C}$, we define the barrier function candidate
\begin{align}
h_{\mathcal{D}}(x,Q) = \sigma(h(x)) \left( \mathcal{D}(x,Q) - \epsilon \right)
\label{eq:CBF_nabla}
\end{align}
where $\epsilon$ is a small positive constant and $\sigma : \mathbb{R} \rightarrow \mathbb{R}$ is a smooth, positive semi-definite function such that: 
(i) $\sigma(0) > 0$ and 
(ii) $\lim_{t \rightarrow \infty} \sigma(t) = 0$. 
The reason for this selection is to ensure that $\mathcal{D}(x,Q) \ge \epsilon > 0$ when $x$ is close to $\partial \mathcal{C}$.

In a similar way than for the CLF \eqref{eq:rotCLF}, the time derivative of 
$h_\mathcal{D}(x,Q)$ is also affine with respect to $u$ and $\omega$. 
This fact allows the corresponding inequality constraint on the dynamics of $h_{\mathcal{D}}(x,Q)$ to be written as an {\it affine} inequality constraint in both $u$ and $\omega$, allowing the use of QPs.
%
%
%

Finally, we propose a modification on the QP-based approach by \cite{Ames2014} to achieve stabilization and safety for \eqref{eq:feedback_system} without the existence of boundary equilibria.
\begin{theorem}
\label{theorem3}
Consider the nonlinear system with dynamics given by \eqref{eq:nonlinear_system} and full rank $g(x)$. Assume a reference, positive definite, non-radial CLF $V_r(x)$, and the CLF $V(x,Q)$ given by \eqref{eq:rotCLF}, where $Q \in \mathcal{SO}(n)$ is virtual state with dynamics given by \eqref{eq:Q_dynamics} and $Q(0) = I_n$. Additionally, assume a convex CBF $h(x)$ and $h_\mathcal{D}(x,Q)$ given by \eqref{eq:CBF_nabla}. The QP
\begin{align}
\mat{ k(x)^\mathsf{T} \!\!\!&\!\!\! \omega(x)^\mathsf{T} \!\!\!&\!\!\! \delta(x) }^\mathsf{T} \!\!= 
\argmin_{\mathclap{\substack{(u,v,w) \in \\ \mathbb{R}^{m+\frac{1}{2}(n^2-n+2)}}}} \, \norm{u}^2 \!+\! q \norm{v}^2 \!+\! p w^2 \label{eq:novel_QP_control} \\
s.t. 
\,\, \dot{V}(x,Q,u,v) + \gamma(V(x,Q)) &\le w \nonumber 
\\
\dot{h}(x,u) + \alpha(h(x)) &\ge 0 \nonumber
\\
\dot{h}_{\mathcal{D}}(x,Q,u,v) + \beta(h_\mathcal{D}(x,Q)) &\ge 0 \nonumber
\end{align}
with $p, q > 0$ and function $\beta \in \mathcal{K}_\infty$ renders the set $\mathcal{C}$ forward invariant and guarantees that no boundary equilibria exist.
%
\end{theorem}

The proof of this result is presented in Appendix \ref{proof3}.
%
%
The first and second constraints on \eqref{eq:novel_QP_control} are the usual CLF and CBF constraints from \eqref{eq:QP_control}, guaranteeing stabilization and safety with respect to set $\mathcal{C}$ as soft and hard constraints, respectively.
The third constraint guarantees that $h_\mathcal{D}(x,Q) \ge 0$ everywhere. Since $\omega(x)$ determines the dynamics of $Q$ (as given by \eqref{eq:Q_dynamics}), its effect is of rotating the reference, non-radial CLF around the origin, ensuring that $\mathcal{D}(x,Q) \ge \epsilon$ when the trajectories are close to the boundary $\partial \mathcal{C}$ (since $\sigma(h(x)) > 0$ in this case).

%


\section{Simulation Results}

In this section, we present numerical examples of our approach for the integrator and different nonlinear systems. We use the same CLF used before as the reference CLF $V_r(x) = 0.5 \lambda_1 x_1^2 + 0.5 \lambda_2 x_2^2$, $\lambda_1 = 6$ and $\lambda_2 = 1$ and CBF $h(x) = 0.5 \norm{x - x_c}^2 - 0.5 r^2$, with $x_c = \mat{ 0 \!\!&\!\! 3 }^\mathsf{T}$ and $r = 1.5$. For the proposed controller on Theorem \ref{theorem3}, we have used $p = q = 5$, $\gamma = \alpha = \beta = 1$ and $\epsilon = 0.1$.

\begin{figure}
\begin{center}
\includegraphics[width=0.86\columnwidth]{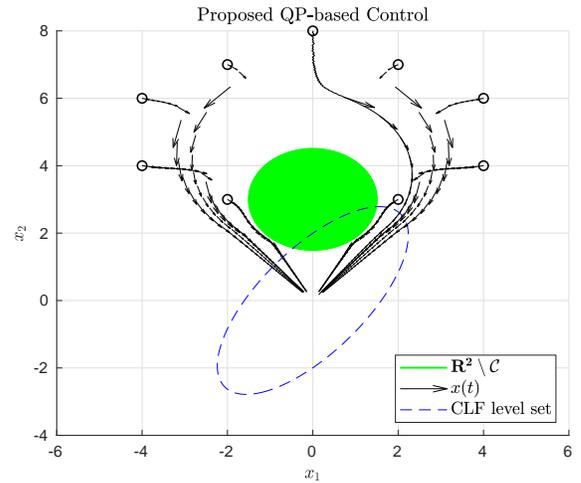}
\end{center}
\vspace{-5mm}
\caption{System trajectories for the closed-loop system with the proposed QP-based control for the integrator.}
\label{fig:novelQP_integrator}
\end{figure}

In the integrator case,  $f(x) = 0$ and $g(x) = I_n$, and \eqref{eq:parallel} simplifies to $\mathcal{D}(x,Q) = \frac{1}{2} \nabla V_r(Qx)^\mathsf{T} Q \mathcal{P}_{\nabla h} Q^\mathsf{T} \nabla V_r(Qx)$. Note that in this case, $\mathcal{D}(x,Q) = 0$ consists of the set of points where the gradients $\nabla V(x,Q) = Q^\mathsf{T} \nabla V_r(Qx)$ and $\nabla h(x)$ are collinear. Figure \ref{fig:novelQP_integrator} shows different system trajectories for the same initial conditions shown in \fref{fig:initial_cond}. 
The red equilibrium point on top of the obstacle does not exist for the closed-loop system with the proposed controller, and all trajectories are attracted the stable origin instead.
We also show the level set of the CLF for a particular state on a particular trajectory, illustrating that $V(x,Q)$ is actually a rotated version of the reference CLF $V_r(x)$. Our proposed control rotates $V_r(x)$ in order to avoid the trajectory to approach the set defined by $\mathcal{D}(x,Q) = 0$.

\begin{figure}
\begin{center}
\includegraphics[width=0.86\columnwidth]{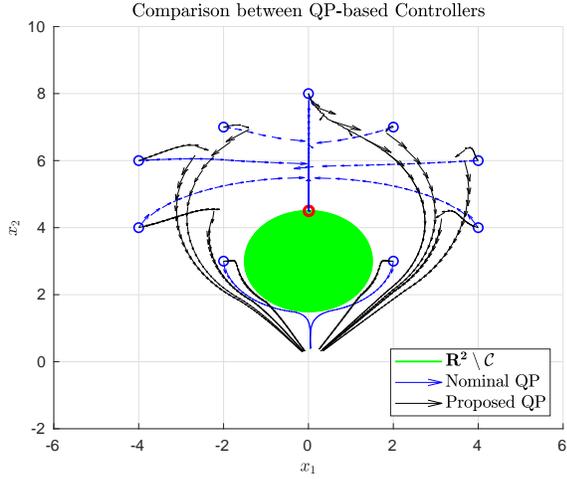}
\end{center}
\vspace{-5mm}
\caption{System trajectories for the closed-loop system with nominal and proposed QP-based controller for $f(x) = f_1(x)$ and $g(x) = I_n$ with a circular obstacle.}
\label{fig:novelQP_nonlinear1}
\end{figure}

\begin{figure}
\begin{center}
\includegraphics[width=0.86\columnwidth]{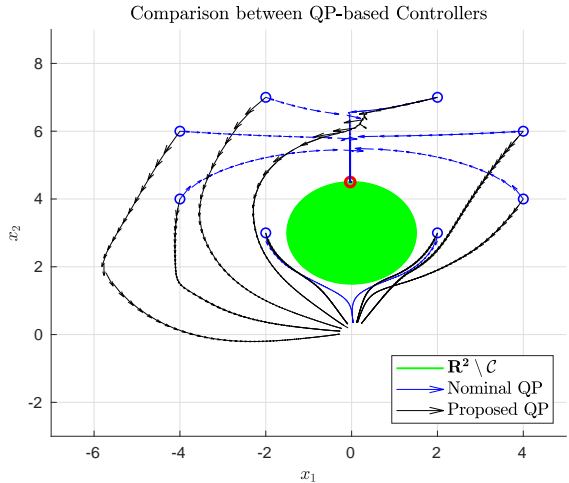}
\end{center}
\vspace{-5mm}
\caption{System trajectories for the closed-loop system with nominal and proposed QP-based controller $f(x) = f_2(x)$ and $g(x) = I_n$ with a circular obstacle.}
\label{fig:novelQP_nonlinear2}
\end{figure}

In Figs. \ref{fig:novelQP_nonlinear1} and  \ref{fig:novelQP_nonlinear2}, we show the results for two different nonlinear systems with $f(x)$ given by $f_1(x) = 0.1 \norm{x} \mat{ 1 \!&\! 1 }^\mathsf{T}$ and $f_2(x) = 0.1(\norm{x} - x^\mathsf{T} x) \mat{ 1 \!&\! 1 }^\mathsf{T}$, respectivelly, both with $g(x) = I_n$. For the same initial conditions, the trajectories show the evolution of the closed-loop system state for both controllers. As before, the nominal QP-based controller proposed by \cite{Ames2014} introduces the same undesired equilibrium point in the closed-loop system, while the trajectories for our proposed controller are attracted towards the origin. This behavior is obtained by the rotation of the CLF around the origin induced by the QP-based controller described in Theorem \ref{theorem3}.

\section{CONCLUSIONS}
In the present work, we have formally demonstrated general conditions for the formation of undesired equilibria on the closed-loop system with the QP-based controller unifying CLFs and CBFs as proposed by \cite{Ames2014}, and showed that these equilibria can be asymptotically stable even for simple systems. Additionally, we have proposed a modification of the controller proposed in \cite{Ames2014} using a strategy that effectivelly avoids the conditions for the formation of boundary equilibria. Future works include the extension of the proposed controller for non-convex obstacles.


\section{APPENDIX}
\label{appendix}

\subsection{Proof of Theorem \ref{theorem1}}
\label{proof1}

The Lagrangian associated to control law \eqref{eq:QP_control} is given by
\begin{align}
\mathcal{L} = \frac{1}{2} \norm{u}^2 + \frac{1}{2} p w^2 +& \lambda_1 ( L_f V + L_g V u + \gamma(V) - w ) \nonumber \\
														      -& \lambda_2 ( L_f V + L_g h \, u + \alpha(h) )
\label{eq:general_lagrangian}
\end{align}
where the dependency of $V(x)$, $L_f V(x)$, $L_g V(x)$, $h(x)$, $L_f h(x)$, $L_g h(x)$ and $u(x)$ on the state was dropped for compactness of notation. Here, 
$\lambda_1, \lambda_2$ are the Karush-Kuhn-Tucker (KKT) multipliers, and the KKT conditions are
\begin{align}
\frac{\partial \mathcal{L}}{\partial u} = u + \lambda_1 L_g V - \lambda_2 L_g h = 0 \label{eq:general_KKT1} \\
\frac{\partial \mathcal{L}}{\partial w} = p w - \lambda_1 = 0 \label{eq:general_KKT2} \\
\lambda_1 ( L_f V + L_g V u + \gamma(V) - w ) = 0 \label{eq:general_KKT3} \\
\lambda_2 ( L_f V + L_g h \,u + \alpha(h) ) = 0 \label{eq:general_KKT4}
\end{align}
with $\lambda_1, \lambda_2 \ge 0$. At this point, we need to distinguish four different cases depending on the activation of each constraint.


{\bf Case 1.} In this case, both constraints are inactive, that is $L_f V + L_g V u + \gamma(V) - w < 0$ and $L_f h + L_g h\,u + \alpha(h) > 0$ and $\lambda_1 = \lambda_2 = 0$. From the KKT conditions, $k(x) = 0$, $\omega(x) = 0$.
%
%
However, since $\gamma(V) < 0$ in this case, this solution never holds.

%


{\bf Case 2.} In this case, only the CLF constraint is active, that is $L_f V + L_g V^\mathsf{T} u + \gamma(V) = \delta$ and $L_f h + L_g h^\mathsf{T} u + \alpha(h) > 0$, and $\lambda_1 \ge 0, \lambda_2 = 0$. Using \eqref{eq:general_KKT1}-\eqref{eq:general_KKT2}, the solution $k(x)$ is
\begin{align}
k(x) = - \frac{L_f V + \gamma(V)}{p^{-1} + \norm{L_g V}^2} L_g V^\mathsf{T}
\label{eq:general_2case}
\end{align}
Since only the CBF is inactive in this case, \eqref{eq:general_2case} holds for ${\bf \Omega_{\overline{cbf}}^{clf}}$ as defined in Theorem \ref{theorem1}.
%
%
From the closed-loop system \eqref{eq:feedback_system}, the equilibrium points are given by
\begin{align}
f_{cl}(x) = f(x) - \frac{L_f V + \gamma(V)}{p^{-1} + \norm{L_g V}^2} G \nabla V = 0
\label{eq:general_equilibrium_case2}
\end{align}
%
%
There are two possible solutions for \eqref{eq:general_equilibrium_case2}:
%
\begin{enumerate}
\item[(i)] $f(x)=0$ and $G(x)\nabla V(x) = 0$ for some $x \in {\bf \Omega_{\overline{cbf}}^{clf}}$.
%
\item[(ii)] $f(x) = \kappa G(x) \nabla V(x)$: substituting this expression into \eqref{eq:general_equilibrium_case2} yields $\kappa = p \gamma(V)$.
%
Then, any  $x \in {\bf \Omega_{\overline{cbf}}^{clf}}$ such that $f(x) = p \gamma(V) G \nabla V$ is an equilibrium point of \eqref{eq:feedback_system}. 
\end{enumerate}
%
Both solutions hold strictly on $\text{int}(C)$. This motivates the definition of the set of {\bf interior equilibria} $\mathcal{E}_{\text{int}}$ on \eqref{eq:interior_equilibria}.


{\bf Case 3.} In this case, only the CBF constraint is active, that is $L_f h + L_g h \, u + \alpha(h) = 0$ and $L_f V + L_g V \, u + \gamma(V) - \delta < 0$, with $\lambda_1 = 0, \lambda_2 \ge 0$. Using the KKT conditions, the corresponding solution for $k(x)$ is given by
\begin{align}
k(x) = - \norm{L_g h}^{-2} \left( L_f h + \alpha(h) \right) L_g h^\mathsf{T}
\nonumber
\end{align}
Since the CLF is inactive in this case, this solution holds for
\begin{align}
&{\bf \Omega_{cbf}^{\overline{clf}}} = \Big\{ x \in \mathbb{R}^n : L_f h + \alpha(h) \le 0 \,, \nonumber \\ 
&L_g V L_g h^\mathsf{T} \left( L_f h + \alpha(h) \right) > \left( L_f V + \gamma(V) \right) \norm{L_g h}^2 \Big\} \nonumber
\end{align}
The equilibrium points are given by $f_{cl}(x) = 0$, yielding
\begin{align}
f_{cl}(x) = f(x) - \norm{L_g h}^{-2} \!\left( L_f h + \alpha(h) \right) G \nabla h = 0
\label{eq:general_equilibrium_case3}
\end{align}
%
%
Similarly to the previous case, the solutions of \eqref{eq:general_equilibrium_case3} are:
\begin{enumerate}
\item[(i)] $f(x)=0$ and $G(x)\nabla h(x) = 0$ for some $x \in {\bf \Omega_{cbf}^{\overline{clf}}}$. 
%
\item[(ii)] $f(x) = \kappa G(x) \nabla h(x)$: substituting this expression into \eqref{eq:general_equilibrium_case3} yields $\alpha(h) = 0$, which means that this condition only happens for $x \in \partial \mathcal{C}$.
\end{enumerate}
However, points satisfying these conditions are not compatible with the first condition of ${\bf \Omega_{cbf}^{\overline{clf}}}$.
Therefore, no equilibria exists in this case.


{\bf Case 4.} This is the case where both constraints are active, that is $L_f V + L_g V^\mathsf{T} u + \gamma(V) - \delta = 0$ and $L_f h + L_g h^\mathsf{T} u + \alpha(h) = 0$. Therefore, $\lambda_1,\lambda_2 \ge 0$ and we have to solve the KKT conditions \eqref{eq:general_KKT1}-\eqref{eq:general_KKT2} simultaneously for both $\lambda_1$ and $\lambda_2$, yielding the following matrix equation:
\begin{align}
\begin{bmatrix}
p^{-1} + \norm{L_g V}^2 \!\!&\!\! -L_g V L_g h^\mathsf{T} \\
L_g V L_g h^\mathsf{T} \!\!&\!\! -\norm{L_g h}^2
\end{bmatrix} \!\!
\begin{bmatrix}
\lambda_1 \\
\lambda_2
\end{bmatrix}
\!\!=\!\!
\begin{bmatrix}
L_f V + \gamma(V) \\
L_f h + \alpha(h)
\end{bmatrix}
\label{eq:general_4case_matrix}
\end{align}
The determinant of the matrix on the left-side of \eqref{eq:general_4case_matrix} is
\begin{align}
\Delta = (L_g V L_g h^\mathsf{T})^2 - ( p^{-1} + \norm{L_g V}^2) \norm{L_g h}^2 \nonumber
\end{align}
Note that $\Delta \le 0$ as long as $p > 0$. Consider the two cases:
\begin{enumerate}
\item[(i)] $\Delta = 0$: in this case, the matrix on the left-side of \eqref{eq:general_4case_matrix} loses rank when $L_g h = 0$, and a solution can only exist for $L_f h + \alpha(h) = 0$. In this case, $k(x)$ is given by \eqref{eq:general_2case}.
\item[(ii)] $\Delta < 0$: in this case, the solution is given by
\begin{align}
k(x) &= -\lambda_1 L_g V^\mathsf{T} +\lambda_2 L_g h^\mathsf{T}
\label{eq:general_4case}
\end{align}
\end{enumerate} 
with $\lambda_1$ and $\lambda_2$ drawn from
\begin{align}
\lambda_1 &=
\frac{1}{\Delta} \left( ( L_f h \!+\! \alpha(h) ) L_g V L_g h^\mathsf{T} - ( L_f V \!+\! \gamma(V) ) \norm{L_g h}^2 \right) \nonumber \\
%
\lambda_2 &= \frac{1}{\Delta} \big( ( L_f h \!+\! \alpha(h) )\left( \norm{L_g V}^2 \!+\! p^{-1} \right) - \label{eq:general_4case_lambda} \\ 
&\qquad \qquad \qquad \qquad \qquad \qquad ( L_f V \!+\! \gamma(V) ) L_g V L_g h^\mathsf{T} \big) \nonumber
\end{align}
%

For both $\Delta = 0$ or $\Delta > 0$, this solution holds for
\begin{align}
{\bf \Omega_{cbf}^{clf}} = \Big\{ &x \in \mathbb{R}^n : L_g V L_g h^\mathsf{T} \left( \frac{L_f h + \alpha(h)}{L_f V + \gamma(V)} \right) \!\le\! \norm{L_g h}^2 \,, \nonumber \\ 
&L_g V L_g h^\mathsf{T} \ge \left( \frac{L_f h \!+\! \alpha(h)}{L_f V \!+\! \gamma(V)} \right) \left(\norm{L_g V}^2 \!+\! p^{-1} \right) \Big\} \nonumber
\end{align}
Using solution \eqref{eq:general_4case} on the closed-loop system \eqref{eq:feedback_system}, the equilibrium condition $f_{cl}(x) = 0$ is given by
\begin{align}
f_{cl}(x) = f(x) - \lambda_1 G \nabla V + \lambda_2 G \nabla h = 0
\label{eq:general_equilibrium_case4}
\end{align}
with $\lambda_1, \lambda_2$ drawn from \eqref{eq:general_4case_lambda} ($\Delta < 0$). In case $\Delta = 0$, $\lambda_2$ is not defined, but the last term of \eqref{eq:general_equilibrium_case4} is zero anyway since $L_g h = 0$. By carefully looking at \eqref{eq:general_equilibrium_case4}, the following general conditions for the occurrence of a valid solution arise.
\begin{enumerate}
\item[(i)] $f(x) = 0$ for all $x \in \mathbb{R}^n$, and $\nabla V(x) || \nabla h(x)$
\item[(ii)] $\nabla h = 0$ or $\nabla h \in \mathcal{N}(G)$, and $f(x) || G(x) \nabla V(x)$
\item[(iii)] $\nabla V = 0$ or $\nabla V \in \mathcal{N}(G)$, and $f(x) || G(x) \nabla h(x)$
\item[(iv)] $\nabla V(x) || \nabla h(x)$, and $f(x) || G(x) \nabla h(x)$
\end{enumerate}
%


Note that solution (ii) is valid for $\Delta = 0$, with solution given by \eqref{eq:general_2case}. In this case, if $f(x) = \kappa G(x) \nabla V(x)$ for some constant $\kappa \in \mathbb{R}$, replacing it on \eqref{eq:general_equilibrium_case4} yields $\kappa = p \gamma(V) \in \mathbb{R}_{> 0}$. Therefore, every $x \in {\bf \Omega_{cbf}^{clf}}$ such that $G \nabla h(x) = 0$ with $f(x) = p \gamma(V(x)) G(x) \nabla V(x)$ is an equilibrium point. 
Since this solution is valid for $L_f \nabla h + \alpha(h) = 0$, and $L_f \nabla h = 0$ on the equilibrium points, we conclude that these equilibria occur on the boundary $h(x) = 0$ of the safe set.
For solutions (iii) and (iv), using a similar reasoning, it is possible to show that these equilibria also can only occur on $\partial \mathcal{C}$. 
Note that any $x \in \mathbb{R}^n$ such that the null space of matrix
$
\begin{bmatrix}
f(x) \!\!&\!\! G(x) \nabla V(x) \!\!&\!\! G(x) \nabla h(x)
\end{bmatrix}
$
is nontrivial satisfies at least one of the conditions (i), (ii), (iii) or (iv). This motivates the definition of the set of {\bf boundary equilibria} $\mathcal{E}_{\partial \mathcal{C}}$ on \eqref{eq:boundary_equilibria}. \qed

\subsection{Proof of Theorem \ref{theorem2}}
\label{proof2}

For the integrator system, the Jacobian matrix of $f_{cl}(x^\star)$ in \eqref{eq:feedback_system} using solution \eqref{eq:general_4case} (valid for a $x^\star \in \mathcal{E}_{\partial \mathcal{C}}$) is given by
\begin{align}
J_{cl}(x^\star)
&\!=\! - \norm{\nabla h}^{-2} \!\left( p \gamma(V) \mathcal{P}_{\nabla h} (H_V \!-\! c H_h) \!+\! \alpha'(0) \nabla h \nabla h^\mathsf{T} \right) \nonumber
\end{align}
%
%
Left-multiplying the eigenvalue equation $J_{cl}(x^\star) v_i = \lambda_i v_i$ by $\nabla h(x^\star)$ and using property (v) of the projection matrix yields $\lambda_i \nabla h^\mathsf{T}(x^\star) v_i = - \alpha'(0) \nabla h^\mathsf{T}(x^\star) v_i$, which shows that all eigenvalues associated to eigenvectors such that $v^\mathsf{T}_i \nabla h(x^\star) \ne 0$ are given by $\lambda_i = -\alpha'(0)$, and therefore are strictly negative (since $\alpha'(0) > 0$).
In this case, the eigenvalue equation can be rewritten as
\begin{align}
p \gamma(V) \mathcal{P}_{\nabla h} \left( H_V \!-\! c H_h \!-\! \frac{\alpha'(0)}{p \gamma(V)} I_n \right) v_i \!=\! 0
\label{eq:eigen_expansion_nonzero}
\end{align} 
Equation \eqref{eq:eigen_expansion_nonzero} shows that the vector on which $\mathcal{P}_{\nabla h}$ operates lies in $\mathcal{N}(\mathcal{P}_{\nabla h})$. However, from property (iii) of the scaled projection matrices, the null space of $\mathcal{P}_{\nabla h}$ is a one dimensional subspace generated by $\nabla h(x^\star)$. Therefore, the corresponding engenvector $v_i$ must be unique, meaning that $\lambda_i = - \alpha'(0)$ is an unique eigenvalue of $J_{cl}(x^\star)$ such that $v_i^\mathsf{T} \nabla h(x^\star) \ne 0$. All remaining $n-1$ eigenvectors $v_1, v_2, \cdots, v_{n-1} \in \mathbb{R}^n$ must lie in the ($n-1$)-dimensional projective hyperplane with normal given by $\nabla h(x^\star)$.
Therefore, the stability of $x^\star \in \mathcal{E}_{\partial \mathcal{C}}$  is completely determined by the $n-1$ eigenvalues of $J_{cl}(x^\star)$ associated to these eigenvectors.
Then, $x^\star \in \mathcal{E}_{\partial \mathcal{C}}$ is asymptotically stable if
\begin{align}
v_i^\mathsf{T} J_{cl}(x^\star) v_i = \lambda_i \norm{v_i}^2 < 0 \,, \quad v_i^\mathsf{T} \nabla h(x) = 0 \nonumber
\end{align}
Using property (iv) of the projection matrix, yields
\begin{align}
\lambda_i \norm{v_i}^2 &= - p \gamma(V) \norm{\nabla h}^{-2} v_i^\mathsf{T} \mathcal{P}_{\nabla h} (H_V(x^\star) - c H_h(x^\star)) v_i \nonumber \\
&= - p \gamma(V) \norm{\nabla h}^{-4} v_i^\mathsf{T} (H_V(x^\star) - c H_h(x^\star)) v_i < 0 \nonumber
\end{align}
Since $p \gamma(V) \norm{\nabla h}^{-4} > 0$, the equilibrium point $x^\star \in \mathcal{E}_{\partial \mathcal{C}}$ is asymptotically stable if $H_V(x^\star) - c H_h(x^\star) > 0$. \qed

\subsection{Proof of Theorem \ref{theorem3}}
\label{proof3}

First, note that the CBF $h(x)$ defining the set $\partial \mathcal{C}$ is a zeroing barrier function \cite{Ames2017}. Therefore, for any $x \in \partial \mathcal{C}$, $\dot{h}(x) \ge -\alpha(h(x)) = 0$. Then, as established by \cite{Ames2017}, the set $\mathcal{C}$ is forward invariant.

Similarly, since $h_\mathcal{D}(x)$ is also a zeroing barrier function, for any $x \in \mathbb{R}^n$ such that $h_{\mathcal{D}}(x) = 0$, $\dot{h}_{\mathcal{D}}(x) \ge -\beta(h_{\mathcal{D}}(x)) = 0$, which establishes that the set defined by $\mathcal{C}_\mathcal{D} = \{x \in \partial \mathcal{C}: \mathcal{D}(x,Q) \ge \epsilon \}$ is also forward invariant.
To prove that no boundary equilibrium exists, first note that the equilibrium conditions are given by $f_{cl}(x) = 0$ and $\omega(x) = 0$, since the system state consists of $\{x,Q\} \in \mathbb{R}^n \times \mathcal{SO}(n)$.
Next, we have to consider all possible solutions for $k(x)$ and $\omega(x)$ imposed by \eqref{eq:novel_QP_control} and show that no solution for these equilibrium conditions is possible on the boundary of $\mathcal{C}$.

%
The solutions for the QP \eqref{eq:novel_QP_control} can be divided into two major groups, depending on the activation of the third constraint.
If the third constraint is inactive, we have $\dot{h}_\mathcal{D} + \beta(h_\mathcal{D}) > 0$ and $\lambda_3 = 0$, and the first equilibrium condition $f_{cl}(x) = 0$ occurs on the same conditions as those discussed in \sref{proof3}. Assume that a boundary equilibrium point $x^\star \in \partial \mathcal{C}$ exists in this case. Then, $\mathcal{D}(x^\star,Q) = 0$, which is a contradiction with the fact that $\mathcal{C}_\mathcal{D}$ is forward invariant.

A similar reasoning can be done by studying the equilibrium conditions resulting from the solutions of the QP \eqref{eq:novel_QP_control} when the third constraint is active. In this case, we have $\dot{h}_\mathcal{D} + \beta(h_\mathcal{D}) = 0$ and $\lambda_3 \ge 0$. Once again, we have to consider all cases, depending on the activation of the first (CLF) and second (CBF) constraints.
Using the KKT conditions to compute the general solutions along with the time derivative of \eqref{eq:CBF_nabla}, and using \eqref{eq:feedback_system}, the general equilibrium conditions holding for all cases are given by
\begin{align}
f \!-\! \lambda_1 G \nabla V \!+\! \bar{\lambda}_2 G \nabla h \!+\! \lambda_3 \sigma(h) G \nabla \mathcal{D} &= 0 \label{eq:novel_equilibrium_case4_1} \\
\lambda_1 \nabla V^\mathsf{T} \mathcal{O}_n(x) - \lambda_3 \sigma(h) \nabla_{\!Q} \mathcal{D} &= 0
\label{eq:novel_equilibrium_case4_2}
\end{align}
where $\bar{\lambda}_2 = \lambda_2 + \lambda_3 \sigma'(h) (\mathcal{D} -\epsilon)$.
The gradients $\nabla \mathcal{D} \in \mathbb{R}^n$ and $\nabla_{\!Q} \mathcal{D} \in \mathbb{R}^{\frac{1}{2}n(n-1)}$ are computed from the time derivative of \eqref{eq:parallel}, and their expressions are:
\begin{align}
\nabla \mathcal{D} &= ( H_V G + \Gamma^\mathsf{T}_{g,\nabla V} ) ( \mathcal{P}_{f} \!+\! \mathcal{P}_{G \nabla h} ) G \nabla V + \nonumber
\\
&\quad \,\,( H_h G + \Gamma^\mathsf{T}_{g,\nabla h} ) \mathcal{P}_{G \nabla V} G \nabla h + \nabla f^\mathsf{T} \mathcal{P}_{G \nabla V} f \nonumber \\
\nabla_{\!Q} \mathcal{D} &= ( H_V \mathcal{O}_n(x) - \mathcal{O}_n(\nabla V) )^\mathsf{T} G ( \mathcal{P}_{f} \!+\! \mathcal{P}_{G \nabla h} ) G \nabla V \nonumber
\end{align}
where the matrix $\Gamma_{g,v} \in \mathbb{R}^{n \times n}$ for $v \in \mathbb{R}^n$ is defined as
\begin{align}
\Gamma_{g,v} = \sum^m_{i=1} ( g_i^\mathsf{T} v I_n + g_i v^\mathsf{T} ) \nabla g_i \nonumber
\end{align}
where the $g_i(x) \in \mathbb{R}^n$ are the columns of $g(x) \in \mathbb{R}^{n \times m}$.

Note that, in general, \eqref{eq:novel_equilibrium_case4_1} holds for $f(x) = \kappa_1 G \nabla V$ and $G \nabla h = \kappa_2 G \nabla V$.
Assume that a boundary equilibrium point $x^\star \in \partial \mathcal{C}$ exists in this case. Then, $\mathcal{D}(x^\star,Q) = 0$, which is a contradiction with the fact that $\mathcal{C}_\mathcal{D}$ is forward invariant. In other cases where $\lambda_1$ or $\lambda_2$ are zero, we conclude that \eqref{eq:novel_equilibrium_case4_1} never holds or the solution is incompatible with the set where the corresponding QP solution is valid, for all $x \in \partial \mathcal{C}$. Therefore, we conclude that no boundary equilibrium can exist on the closed-loop system.

\qed



\bibliographystyle{plain}
\bibliography{references}

\end{document}